\documentclass [aps,prc,superscriptaddress,reprint]{revtex4-1}
\usepackage{graphicx} 
\usepackage{epstopdf}
\usepackage{dcolumn} 
\usepackage{bm}
\usepackage{hyperref}
\usepackage{array}
\usepackage{setspace}
\usepackage{booktabs}
\hypersetup{
    colorlinks=true,
    linkcolor=blue,
    filecolor=gray,
    urlcolor=blue,
    citecolor=blue,
}
\begin{document}
\title{Influence of entrance channel on production cross section of exotic actinides in multinucleon transfer reactions}
\author{Peng-Hui Chen}
\email{Corresponding author: chenpenghui@yzu.edu.cn}
\affiliation{College of Physics Science and Technology, Yangzhou University, Yangzhou 225009, China}

\author{Geng Chang}
\affiliation{College of Physics Science and Technology, Yangzhou University, Yangzhou 225009, China}

\author{Xiang-Hua Zeng}
\affiliation{College of Physics Science and Technology, Yangzhou University, Yangzhou 225009, China}
\affiliation{College of Electrical, Power and Energy Engineering, Yangzhou University, Yangzhou 225009, China }

\author{Zhao-Qing Feng}
\email{Corresponding author: fengzhq@scut.edu.cn}
\affiliation{School of Physics and Optoelectronics, South China University of Technology, Guangzhou 510641, China}

\date{\today}
\begin{abstract}
 Within the framework of the dinuclear system model, the influence of mass asymmetry and the isospin effect on the production of exotic actinides have been investigated systematically. The isotopic yields populate in multinucleon transfer reactions of $^{48}$Ca, $^{86}$Kr, $^{136}$Xe, $^{238}$U bombarding on $^{248}$Cm are analyzed and compared to the available experimental data. Systematics on the production of unknown actinides from Ac to Lr via the available stable elements on the earth (from Ar to U) as projectiles-induced reactions with $^{232}$Th, $^{238}$U and $^{248}$Cm are investigated thoroughly. Potential energy surface and total kinetic energy distribution for the reaction system are calculated and can be used to predict the production cross-section trends. It is found that the heavier projectile leads to the wider isotopic chain distribution for the same target. The heavier target-based reactions prefer to produce plenty of exotic actinides through both mechanisms of deep-inelastic and quasi-fission reactions. Isospin relaxation plays a crucial role in the colliding process, resulting in actinide isotopic distribution tends to shift to the drip lines. Massive new actinides have been predicted at the level of nanonbarn to millibarn. The optimal projectile-target combinations and beam energies were proposed for the forthcoming experiments.

\begin{description}
\item[PACS number(s)]
25.70.Jj, 24.10.-i, 25.60.Pj
\end{description}
\end{abstract}

\maketitle

\section{Introduction}

In past decades, a number of new exotic actinides have been identified in laboratories all over the world. There are fifteen elements in the actinide region with the charge number Z = 89-103. In the actinide range, practically these nuclides were produced by such methods: heavy-ion fusion-evaporation reactions (FE), projectile fragmentation (PF), light-particle reactions (LP), neutron capture reactions (NC), heavy-ion transfer reactions (TR), deep-inelastic(DI), thermonuclear tests (TNT) and radioactive decay (RD). Up to now, the most neutron-deficient isotopes for each actinides from Ac to Lr are $^{205}$Ac, $^{207}$Th, $^{211}$Pa, $^{214}$U, $^{219}$Np, $^{228}$Pu, $^{223}$Am, $^{233}$Cm, $^{233}$Bk, $^{237}$Cf, $^{240}$Es, $^{241}$Fm, $^{244}$Md, $^{250}$No and $^{253}$Lr, respectively. Among them, $^{223}$Am and $^{233}$Bk have been produced in multinucleon transfer reactions $^{48}$Ca + $^{248}$Cm. Apart from that, the rest of them have been synthesized by heavy-ion fusion-evaporation reactions. Hitherto, the most neutron-rich isotope for each actinides from Ac to Lr are $^{236}$Ac, $^{238}$Th, $^{239}$Pa, $^{242}$U, $^{244}$Np, $^{247}$Pu, $^{247}$Am, $^{251}$Cm, $^{251}$Bk, $^{256}$Cf, $^{257}$Es, $^{259}$Fm, $^{260}$Md, $^{260}$No and $^{266}$Lr, respectively. Among them, $^{238}$Th, $^{239}$Pa, $^{244}$Np, $^{260}$Md and $^{260}$No were produced through multinucleon transfer  or deep-inelastic reactions. $^{236}$Ac was produced in projectile fragmentation (PF) reaction. PF reactions products are still not so close to the drip line. $^{242}$U, $^{247}$Am, $^{256}$Cf and $^{259}$Fm were created via light-particles reactions. $^{247}$Pu, $^{251}$Cm and $^{257}$Es were produced in neutron capture reactions. $^{260}$No was identified in alpha decay chain based fusion-evaporation reactions. The synthesis information of the most neutron-rich and proton-rich actinide isotopes are listed in Table \ref{tab1}, including the based method, laboratory, country, and year.

Above all, we can see that FE is still the most promising method to create new nuclides nearby the proton drip line. Furthermore, MNT reaction also may approach the proton drip line, particularly after very proton-rich actinides $^{223}$Am and $^{233}$Bk were identified by Devaraja et al. at GSI. Actually, MNT reactions have been used to synthesize very neutron-rich actinides, e.g., $^{238}$Th in $^{18}$O + $^{238}$U with incident energy $60A$ MeV, $^{239}$Pa in $^{18}$O + $^{238}$U with incident energy $50A$ MeV, $^{244}$Np in $^{136}$Xe + $^{244}$Pu with incident energy 835 MeV, based on quasi-fission mechanism (nucleons transfer from target to projectile). $^{260}$Md and $^{260}$No were discovered in MNT reactions of $^{22}$Ne + $^{254}$Es, $^{18}$O + $^{244}$Pu, respectively, based on deep-inelastic mechanism (nucleons transfer from projectile to target). 
Compare to the synthesis of proton-rich actinides, the neutron-rich ones are more difficult to produce and identify because of the $\beta ^-$ decay, which is hardly distinguished and identified by the data capture and detection system state of the art. There are limitations of available projectile-target material in mechanisms of FE, LP, and PF reactions. With particle identification technique development, MNT reactions might be considered the only method that could be used to produce very neutron-rich actinides, especially for the transuranium elements. 
It was well known that the MNT reactions have such advantages that fragment with a wide mass region owing to broad excitation functions in the transfer process. Therefore, more insightful theoretical and experimental studies of the reaction mechanism are required to make precise predictions for the probability of MNT products in such reactions. 

Since the 1970s, MNT reactions of actinide-based targets with projectiles O, Ca, Kr, Xe, U, and Cm have been used to synthesize massive unknown actinides with available cross-section, in the laboratories GSI\cite{PhysRevC.88.054615} and Berkely\cite{PhysRevC.31.1763,PhysRevC.33.1315}.
Based on these available experimental data, it motivates us to explore the influence of mass asymmetry and isospin effect on the production cross-section of exotic actinides in MNT reactions.

In this work, the selection of projectiles as Ar to U induced MNT reactions with the combination of targets $^{232}$Th, $^{238}$U, $^{248}$Cm are calculated with the DNS model.
The aim of this paper is to investigate the entrance channel effect and isospin relaxation on the production cross-sections of actinide products in actinides-based MNT reactions.
The article is organized as follows: In Sec. \ref{sec2} we give a brief description of the DNS model. Calculated results and discussions are presented in Sec. \ref{sec3}. Summary is concluded in Sec. \ref{sec4}.

\begin{table}[h!]
\begin{ruledtabular}
\caption{The most proton- and neutron-rich isotopes for each actinide element were listed in the table, which includes their production of the method, laboratory, country, and year. 
FE for heavy-ion fusion-evaporation reaction, LP for light-particle reactions, NC for neutron capture reactions, TR for heavy-ion transfer reactions, PF for projectile fragmentation, and DI for deep-inelastic reactions.}
\label{tab1}
\begin{spacing}{1.19}
\begin{tabular}{cccccc}
\specialrule{0em}{2pt}{2pt}
  Z      & Isotope     & Method       & Laboratory    & Country   & Year  \\
\hline
Ac & $^{205}$Ac  &     FE\cite{zha14}       & Lanzhou       & China     & 2014    \\
Z=89     & $^{236}$Ac  &     PF\cite{CHEN2010234}       & Darmstadt     & Germany   & 2010    \\
\hline
Z=90 & $^{207}$Th  &     FE\cite{PhysRevC.105.L051302}       & Lanzhou       & China     & 2022    \\
Th & $^{238}$Th  &     TR\cite{PhysRevC.59.520}       & Lanzhou       & China     & 1995    \\
\hline
Z=91 & $^{211}$Pa  &     FE\cite{PhysRevC.102.034305}       & Helsinki      & Finland   & 2020   \\
Pa & $^{239}$Pa  &     TR\cite{Yuan1995ANI}       & Lanzhou       & China     & 1995   \\
\hline
Z=92 & $^{214}$U   &     FE\cite{PhysRevLett.126.152502}       & Lanzhou       & China     & 2021    \\
U & $^{242}$U   &     PL\cite{PhysRevC.19.2332}       & Brookhaven    & USA       & 1979    \\
\hline
Z=93 & $^{219}$Np  &     FE\cite{YANG2018212}       & Lanzhou       & China     & 2018    \\
Np & $^{244}$Np  &     TR\cite{article}       & Darmstadt     & Germany   & 1987   \\
\hline
Z=94 & $^{228}$Pu  &     FE\cite{Andreyev1994NewN}       & Dubna         & Russia    & 1994   \\
Pu & $^{247}$Pu  &     NC\cite{FRY201396}       & Dimitrovgrad  & Russia    & 1983   \\
\hline
Z=95 & $^{223}$Am  &     TR\cite{DEVARAJA2015199}       & Darmstadt     & Germany   & 2015    \\
Am & $^{247}$Am  &     LP\cite{PhysRevLett.19.128}       & Los Alamos    & USA       & 1967    \\
\hline
Z=96 & $^{233}$Cm  &     FE\cite{2010The}       & Darmstadt     & Germany   & 2010  \\
Cm & $^{251}$Cm  &     NC\cite{LOUGHEED19781865}       & Livermore     & USA       & 1978   \\
\hline
Z=97 & $^{237}$Bk  &     TR\cite{DEVARAJA2015199}       & Darmstadt     & Germany   & 2015   \\
Bk & $^{251}$Bk  &     NC\cite{DIAMOND1967601}       & Argonne       & USA       & 1967    \\
\hline
Z=98 & $^{237}$Cf  &     FE\cite{1995Spontaneous}       & Dubna         & Russia    & 1995    \\
Cf & $^{256}$Cf  &     LP\cite{PhysRevC.21.972}       & Los Alamos    & USA       & 1980    \\ 
\hline
Z=99 & $^{240}$Es  &     FE\cite{KONKI2017265}       & Darmstadt     & Germany   & 2017    \\
Es & $^{257}$Es  &     NC\cite{LOUGHEED19812239}       & Dimitrovgrad  & Russia    & 1987    \\
\hline
Z=100 &$^{241}$Fm  &     FE\cite{refId0}       & Darmstadt     & Germany   & 2008    \\
Fm  &$^{259}$Fm  &     LP\cite{PhysRevC.21.966}       & Los Alamos    & USA       & 1980    \\
\hline
Z=101 &$^{244}$Md  &     FE\cite{PhysRevLett.124.252502}       & Berkeley      & USA       & 2020    \\
Md &$^{260}$Md  &     DI\cite{PhysRevC.40.770}       & Berkeley      & USA       & 1989    \\
\hline
Z=102 &$^{250}$No  &     FE\cite{PhysRevC.64.054606}       & Dubna         & Russia    & 2003    \\
No &$^{260}$No  &     DI\cite{PhysRevC.31.1801}       & Berkeley      & USA       & 1985    \\
\hline
Lr &$^{253}$Lr  &     FE\cite{2001Decay}       & Darmstadt     & Germany   & 2001    \\
Z=103 &$^{266}$Lr  &     FE\cite{PhysRevLett.112.172501}       & Dubna         & Russia    & 2014    \\
  \end{tabular}
\end{spacing}
\end{ruledtabular}
\end{table}

\section{Model description}\label{sec2}

The cross-sections of the survival fragments produced in MNT reactions and fusion-evaporation residue cross-sections are evaluated
\begin{eqnarray}\label{mnt}
&&\sigma_{\rm sur}(Z_{1},N_{1},E_{\rm c.m.})= 
\sum_{J=0}^{J_{\max}} \sigma_{cap} \int f(B) \nonumber \\
&& \times  P(Z_{1},N_{1},E_{1},J_{1},B) \times W_{\rm sur}(E_{1},J_{1},s) dB. 
\end{eqnarray}
The equation for computing the cross-section of primary fragments were the above formula without $W_{\rm sur}(E_{1},J_{1},s)$.
The $J_{\rm max}$ is the maximal angular momentum at grazing collisions. The capture cross-section is evaluated by
$\sigma_{\rm cap} = \pi\hbar^2(2J+1)/(2\mu E_{c.m.}) T(E_{\rm c.m.},J)$
The $T(E_{\rm c.m.},J)$ is the transmission probability of the projectile-target nuclei overcoming the Coulomb barrier to form a composite system and calculated by the Hill-Wheeler formula in Refs.\cite{PhysRevC.76.044606}. The $\mu$ is the reduced mass of relative motion. 

The dynamical multinucleon transfer reactions are described as a diffusion process, in which the resulting distribution probability is obtained by solving a set of master equations numerically in the potential energy surface of the DNS. The time evolution of the distribution probability $P(Z_{1},N_{1},E_{1},\beta, t)$ for fragment 1 with proton number $Z_{1}$, neutron number $N_{1}$, excitation energy $E_{1}$, quadrupole deformation $\beta$ are described by the following master equations:
\begin{eqnarray}
\label{mst}
&&\frac{d P(Z_1,N_1,E_1,\beta,t)}{d t} =  \nonumber \\ 
&&  \sum \limits_{Z^{'}_1}  W_{Z_1,N_1,\beta;Z'_1,N_1,\beta}(t) [d_{Z_1,N_1} P(Z'_1,N_1,E'_1,\beta,t) \nonumber   \\
&& - d_{Z'_1,N_1}P(Z_1,N_1,E_1,\beta,t)]   \nonumber   \\
&& + \sum \limits_{N'_1}  W_{Z_1,N_1,\beta;Z_1,N'_1,\beta}(t) [d_{Z_1,N_1}P(Z_1,N'_1,E'_1,\beta,t) \nonumber   \\ 
&& - d_{Z_1,N'_1}P(Z_1,N_1,E_1,\beta,t)]
\end{eqnarray} 
The $W_{Z_{1},N_{1},\beta;Z^{'}_{1},N_{1},\beta}$($W_{Z_{1},N_{1},\beta,;Z_{1},N^{'}_{1},\beta}$) is the mean transition probability from the channel($Z_{1},N_{1},E_{1},\beta$) to ($Z^{'}_{1},N_{1},E^{'}_{1},\beta$), [or ($Z_{1},N_{1},E_{1},\beta$) to ($Z_{1},N^{'}_{1},E^{'}_{1},\beta$)], and $d_{Z_{1},Z_{1}}$ denotes the microscopic dimension corresponding to the macroscopic state ($Z_{1},N_{1},E_{1}$). The sum is taken over all possible proton and neutron numbers that fragment ($Z^{'}_{1}$,$N^{'}_{1}$) may take, but only one nucleon transfer is considered in the model with the relations $Z^{'}_{1}=Z_{1}\pm1$ and $N^{'}_{1}=N_{1}\pm1$. 
The transfer probability is smoothed with the barrier distribution, which is taken as the asymmetry Gaussian form of $f(B)=\frac{1}{N}\exp\left[-((B-B_{\rm m})/\Delta)^{2}\right]$ with the normalization constant satisfying the unity relation $\int  f(B) dB=1$. The quantities $B_{\rm m}$ and $\Delta$ are evaluated by $B_{\rm m}=(B_{\rm C}+B_{\rm S})/2$ and $\Delta=(B_{\rm C}-B_{\rm S})/2$, respectively. The $B_{\rm C}$ and $B_{\rm S}$ are the Coulomb barrier at waist-to-waist orientation and the minimum barrier with varying the quadrupole deformation parameters of colliding partners.

The motion of nucleons in the interacting potential is governed by the single-particle Hamiltonian. The excited DNS opens a valence space in which the valence nucleons have a symmetrical distribution around the Fermi surface. Only the particles at the states within the valence space are actively for nucleon transfer. The transition probability is related to the local excitation energy and nucleon transfer, which is microscopically derived from the interaction potential in valence space as described in \cite{Chen2016,No75}.

\begin{eqnarray}
\label{trw}
&&W_{Z_{1},N_{1},\beta;Z_{1}^{\prime},N_{1},\beta^{\prime}} = \frac{\tau_{mem}(Z_{1},N_{1},\beta,E_{1};Z_{1}^{\prime},N_{1} \beta ^{\prime},E_{1}^{\prime})}{d_{Z_{1},N_{1}} d_{Z_{1}^{\prime},N_{1}}\hbar^{2}}    \nonumber \\
&& \times \sum_{ii^{\prime}}|\langle  Z_{1}^{\prime},N_{1},E_{1}^{\prime},i^{\prime}|V|Z_{1},N_{1},E_{1},i \rangle|^{2}.
\end{eqnarray}
The memory time $\tau_{\rm mem}$ and interaction element $V$ can be seen in Ref. \cite{Chen2016}. A similar approach is used for the neutron transition coefficient.

The local excitation energy is determined by the dissipation energy from the relative motion and the potential energy surface of the DNS as
\begin{eqnarray}
\label{lee}
\varepsilon^{\ast}(t)=E^{\rm diss}(t)-\left(U(\{\alpha, t\})-U(\{\alpha_{\rm EN}\})\right).
\end{eqnarray}
\begin{eqnarray}
\label{dis}
E^{diss}(t) = E_{c.m.} - B - \frac{<J(t)>[<J(t)>+1]\hbar ^2}{2\zeta}     \nonumber
\\ - <E_{rad}(J,t)>
\end{eqnarray}
The entrance channel quantities $\{\alpha_{\rm EN}\}$ include the proton and neutron numbers, quadrupole deformation parameters and orientation angles being $Z_{ \rm P}$, $N_{\rm P}$, $Z_{\rm T}$, $N_{\rm T}$, $R$, $\beta_{\rm P}$, $\beta_{\rm T}$, $\theta_{\rm P}$, $\theta_{\rm T}$ for projectile and target nuclei with the symbols of $P$ and $T$, respectively. The symbol ${\alpha}$ denotes the quantities of fragments of $Z_{1}$, $N_{1}$, $Z_{2}$, $N_{2}$, $R$, $\beta_{1}$, $\beta_{2}$, $\theta_{1}$, $\theta_{2}$ .
The interaction time $\tau_{\rm int}$ is obtained from the deflection function method \cite{Wo78}. The energy dissipated into the DNS increases exponentially \cite{PhysRevC.76.044606}.

The potential energy surface (PES) of the DNS is evaluated by
\begin{eqnarray}\label{dri}
U_{\rm dr}(t) = Q_{\rm gg}+V_{\rm C}(Z_1,N_1;\beta_1,Z_2,N_2,\beta_2,t)     \nonumber   \\ 
+ V_{\rm N}(Z_1,N_1,\beta_1;Z_2,N_2,\beta_2,t) + V_{\rm def}(t)
\end{eqnarray}
with
\begin{eqnarray}\label{vcn}
V_{\rm def}(t) = \frac{1}{2} C_1 (\beta_1 - \beta' _T (t) )^2 + \frac{1}{2} C_2 (\beta_2 - \beta' _P (t) )^2      \\
C_i = (\lambda-1) { (\lambda+2) R^2_{\rm N} \delta - \frac{3}{2\pi}} \frac{Z^2e^2}{R_{\rm N}(2\lambda+1)}.
\end{eqnarray}
Where, the $Q_{\rm gg}$ derived by the negative binding energies of the fragments $(Z_{\rm i},N_{\rm i})$ were calculated by liquid drop model plus shell correction\cite{MOLLER20161}. The $\theta_{i}$ denotes the angles between the collision orientations and the symmetry axes of the deformed nuclei. $V_{\rm C}$ and $V_{\rm N}$ were calculated by the Wong formular\cite{PhysRevLett.31.766} and double-folding potential\cite{PhysRevC.69.024610}, respectively.
$V_{\rm def}(t)$ is the deformation energy of DNS at the reaction time $t$.
The evolutions of quadrupole deformations of projectile-like and target-like fragments undergo from the initial configuration as
\begin{eqnarray}\label{qde}
\beta' _{\rm T} (t) = \beta _ {\rm T} \exp(-t/\tau_{\rm \beta}) + \beta_1 [ 1 - \exp(-t/\tau_{\rm \beta})],      \nonumber \\
\beta' _{\rm P} (t) = \beta _ {\rm P} \exp{(-t/\tau_{\rm \beta})} + \beta_2 [ 1 - \exp(-t/\tau_{\rm \beta})]
\end{eqnarray}
with the deformation relaxation is $\tau_{\rm \beta}=4\times10^{-21} \ s$.

The total kinetic energy (TKE) of the primary fragment is evaluated by
\begin{equation}\label{tke}
 TKE (A_{1}) = E_{\rm c.m.} + Q_{ \rm gg}(A_{1}) - E^{\rm diss}(A_{1}),
\end{equation}
where $Q_{\rm gg} = M_{\rm P} + M_{\rm T} - M_{\rm PLF} -M_{\rm TLF}$ and $E_{\rm c.m.}$ being the incident energy in the center of mass frame. The mass $M_{\rm P}$, $M_{\rm T}$, $M_{\rm PLF}$ and $M_{\rm TLF}$ correspond to projectile, target, projectile-like fragment and target-like fragment, respectively.
The $E_{1}$ and $J_{1}$ are the excitation energy and the angular momentum for the fragment (Z$_{1}$, N$_{1}$). The survival probability $W_{\rm sur}$ of each fragment is evaluated with a statistical approach based on the Weisskopf evaporation theory \cite{Chen_2016}, in which the excited primary fragments are cooled in evaporation channels $s(Z_{s}, N_{s})$ by $\gamma$-rays, light particles (neutrons, protons, $\alpha$, etc) in competition with the binary fission.

\section{Results and discussion}\label{sec3}

\begin{figure}[htb]
\includegraphics[width=1.\linewidth]{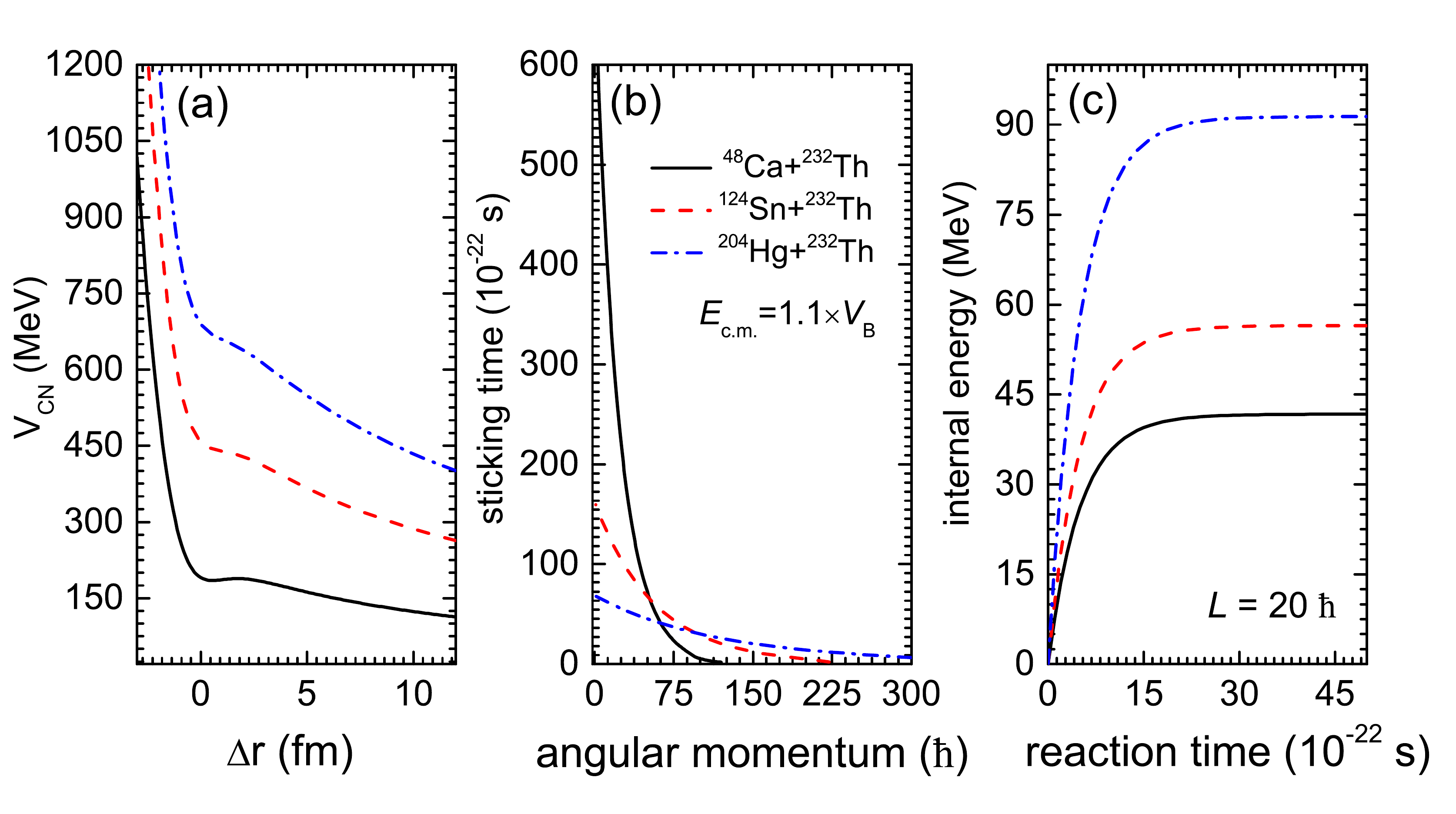}
\caption{\label{fig1}
Solid black, dash red and dash dot blue lines stand for reactions of $^{48}$Ca+$^{232}$Th, $^{124}$Sn+$^{232}$Th and $^{204}$Hg+$^{232}$Th, respectively. These lines indicate interaction potential of the  head-on collisions as a function of surface distance for these three reactions in panel (a), respectively; In panel (b), these lines are the sticking time distribution to angular momentum; These lines are the internal excitation energy distributions of $^{48}$Ca, $^{124}$Sn, $^{204}$Hg on $^{232}$Th to reaction time at $E_{\rm c.m.}$ = 1.1*$V_{\rm B}$ for given impact parameter $L$ = 20 $\hbar$ in panel (c).}
\end{figure}

In heavy-ion damping collisions, colliding partners overcome the Coulomb barrier, and kinetic energy of relative motion transforms rapidly dissipates into internal excitation of a dinuclear system at contact configuration.
The interaction potential distribution to distance, sticking time to impact parameter and internal excitation energy to reaction time for three different mass asymmetry systems of $^{48}$Ca + $^{232}$Th ($\eta$ = 0.171), $^{124}$Sn + $^{232}$Th ($\eta$ = 0.348) and $^{204}$Hg + $^{232}$Th ($\eta$ = 0.468) at incident energy $E_{\rm c.m.}$ = 1.1$\times V_{\rm B}$ are presented in Fig.\ref{fig1}. The interaction potential is a combination of Coulomb potential and nuclide-nuclide potential, which is calculated as a function of the nuclear surface distance between two heavy partners. 
In panel (a), One can see that Coulomb barriers of head-on collision systems are about 185 MeV, 445 MeV and 650 MeV, corresponding to $^{48}$Ca + $^{232}$Th (solid black line), $^{124}$Sn + $^{232}$Th (dash red line) and $^{204}$Hg + $^{232}$Th (dash-dot blue line), respectively. The potential pocket locates nearby contact points almost. 

The sticking time of the collision system is derived from the deflection function, shown as a function of angulmomentum (impact parameter), which decreases exponentially with increasing angular momentum, shown in panel (b). Generally, the mass asymmetry of collision systems is smaller, and their sticking time is longer, due to the small mass asymmetry system having smaller potential energy and potential pocket. 
The internal excitation energy dissipating in the dinuclear system increases exponentially with increasing evolution time, as shown in panel (c). 
The existence of the pocket in the entrance channel is crucial for the compound nucleus formation in fusion reactions, which is the input physical quantity in calculating capture cross-section. 
According to Fig. \ref{fig1}, it was found that there are few MeV potential pockets for the heavy systems, because of the strong Coulomb repulsion between two colliding partners with $Z_1 Z_2 = 1860 $. The lighter collision systems own the deeper potential pocket relatively. The deeper potential pocket collision system led to a longer reaction time relatively.    

\begin{figure}[htb]
\includegraphics[width=1.\linewidth]{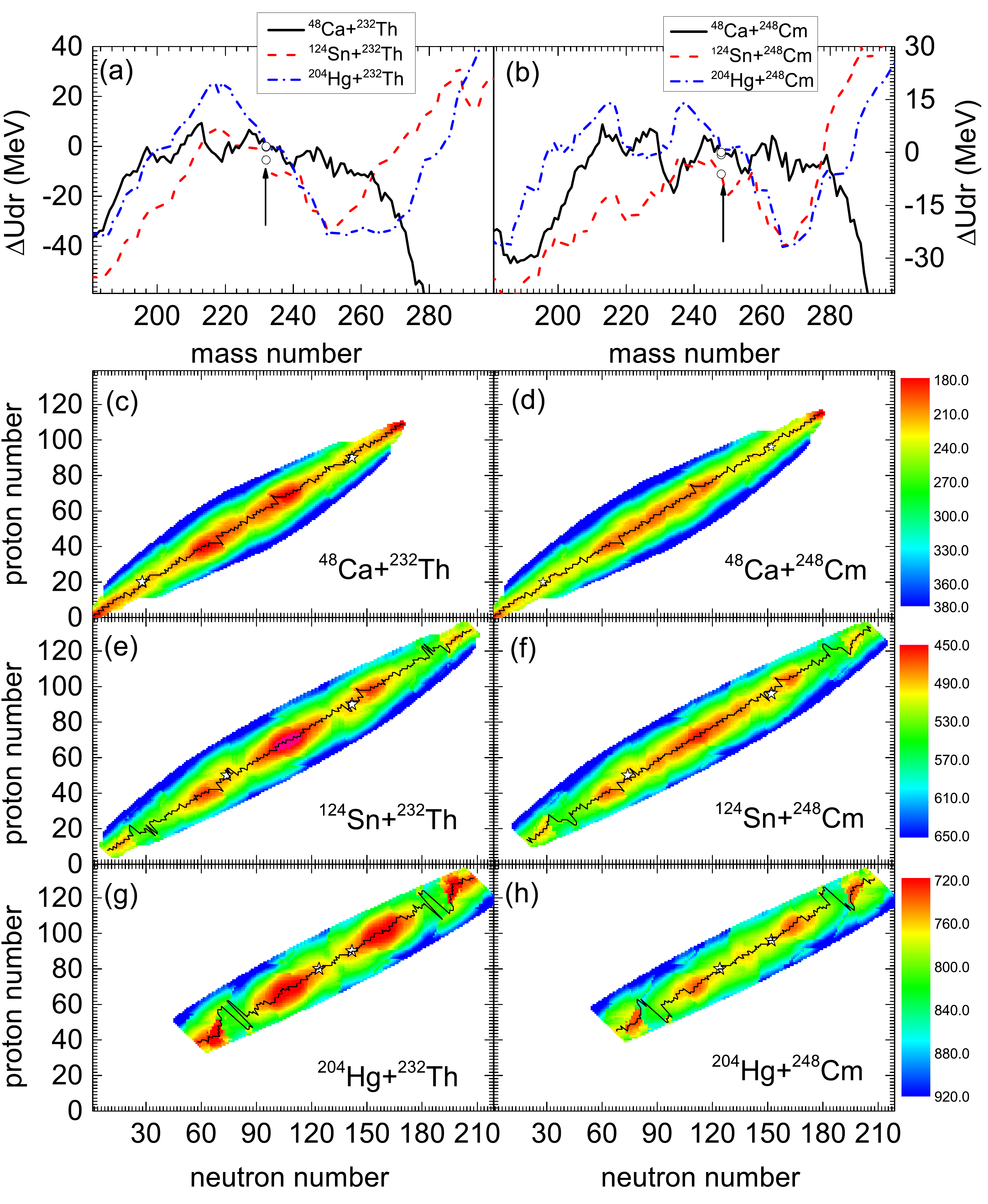}
\caption{\label{fig2} (Color online) In panels (a) and (b), comparison of the driving potentials of $^{48}$Ca, $^{124}$Sn and $^{204}$Hg on $^{248}$Th, $^{248}$Cm, where solid black, dash red and dash-dot blue lines corresponding to $^{48}$Ca, $^{124}$Sn and $^{204}$Hg induced reactions. In panels (c), (e) and (g), potential energy surfaces (PES) of $^{48}$Ca, $^{124}$Sn, $^{204}$Hg on $^{248}$Th, instead of target $^{248}$Th in panels (d), (f) and (h). The open stars indicate the injection points. The solid black lines are minimum value trajectories in two-dimensions potential energy surfaces.}
\end{figure}

Nucleons are transferred between the collision partners resulting in the internal degree of freedom characterizing the nuclear states encountering a rapid rearrangement along the valley line in potential energy surface (PES) as well as dissipating kinetic energy and angular momentum. The calculation of multi-dimensional adiabatic PES for the heavy nuclear system is a quite complex physical problem, until now it is still an open problem. PES and driving potential of head-on collisions of $^{48}$Ca, $^{124}$Sn and $^{204}$Hg bombardment on $^{232}$Th and $^{248}$Cm are calculated by Eq. (\ref{dri}) as a diabetic type with the fixed surface distance, shown in Fig. \ref{fig2}.
The solid black, dash red, dash-dot blue lines, open circles are the valley line in PES of $^{48}$Ca, $^{124}$Sn and $^{204}$Hg induced reactions with $^{232}$Th and $^{248}$Cm, injection points in Fig. \ref{fig2} (a), (b), respectively. 
The valley trajectories in PES are shown as a function of proton number and neutron number, shown in Fig. \ref{fig2}(c), (d), (e), (f), (g), (h).
In order to make comparsion for collision systems with different mass asymmetry, we use $\Delta U_{\rm dr}(Z,N)$ = $U_{\rm dr}(Z,N)$ - $U_{ \rm dr}(Z_ {\rm T},N_ {\rm T}$) to represent the driving potential change trend.
For $^{48}$Ca + $^{232}$Th and $^{48}$Ca + $^{248}$Cm, the pockets locate at $A_{\rm 1}$ = 240 $(Z_{\rm 1} = 94, N_{\rm 1} = 146; Z_{\rm 2} = 42, N_{\rm 2} = 64)$, at $A_{\rm 1} = 254 (Z_{\rm 1} = 100, N_{\rm 1} = 156; Z_{\rm 2} = 16, N_{\rm 2} = 24)$, targets pick up four protons and four neutrons. 
For $^{124}$Sn + $^{232}$Th, a pocket locates at $A_{\rm 1} = 250 (Z_{\rm 1} = 98, N_{\rm 1} = 152; Z_{\rm 2} = 42, N_{\rm 2} = 64)$, target gets four protons and four neutrons. For $^{204}$Hg + $^{232}$Th, the pocket locates at $A_{\rm 1} = 250 - 266 (Z_{\rm 1} = 98 - 104, N_{\rm 1} = 152 - 162; Z_{\rm 2} = 72 - 66, N_{\rm 2} = 114 - 104)$, target gets 8-14 protons and 10-20 neutrons. 
For $^{124}$Sn + $^{248}$Cm, two pockets locate at $A_{\rm 1} = 250 (Z_{\rm 1} = 98, N_{\rm 1} = 152; Z_{\rm 2} = 42, N_{\rm 2} = 64)$ and $= 266 (Z_{\rm 1} = 104, N_{\rm 1} = 162; Z_{\rm 2} = 36, N_{\rm 2} = 54)$.
For $^{204}$Hg + $^{232}$Cm, the pocket appears around $A_{\rm 1} = 266 (Z_{\rm 1} = 104, N_{\rm 1} = 162; Z_{\rm 2} = 66, N_{\rm 2} = 104)$, target picks up fourteen protons and twenty neutrons. 
Neutron subshell numbers (N = 152, 162) play a crucial role in PES.
The DNS fragments towards mass asymmetric location release negative energy, which combined with interaction potential contribute to nucleons transfer. 
From a purely Coulomb perspective, the removal of protons from the target to the projectile is unfavorable.
The spectra exhibit a gaussian-like distribution for potential value belonging to each atomic number. The valley trajectory in the PES is close to the $\beta$-stability line. 
Therefore, we solve the master equations with the PES to get the all-formed fragments probability for these collision systems with different mass asymmetry and dynamical quadrupole deformations. 
From these PES, we can roughly predict the trend of fragments probability diffusion.

\begin{figure}[htb]
\includegraphics[width=1.\linewidth]{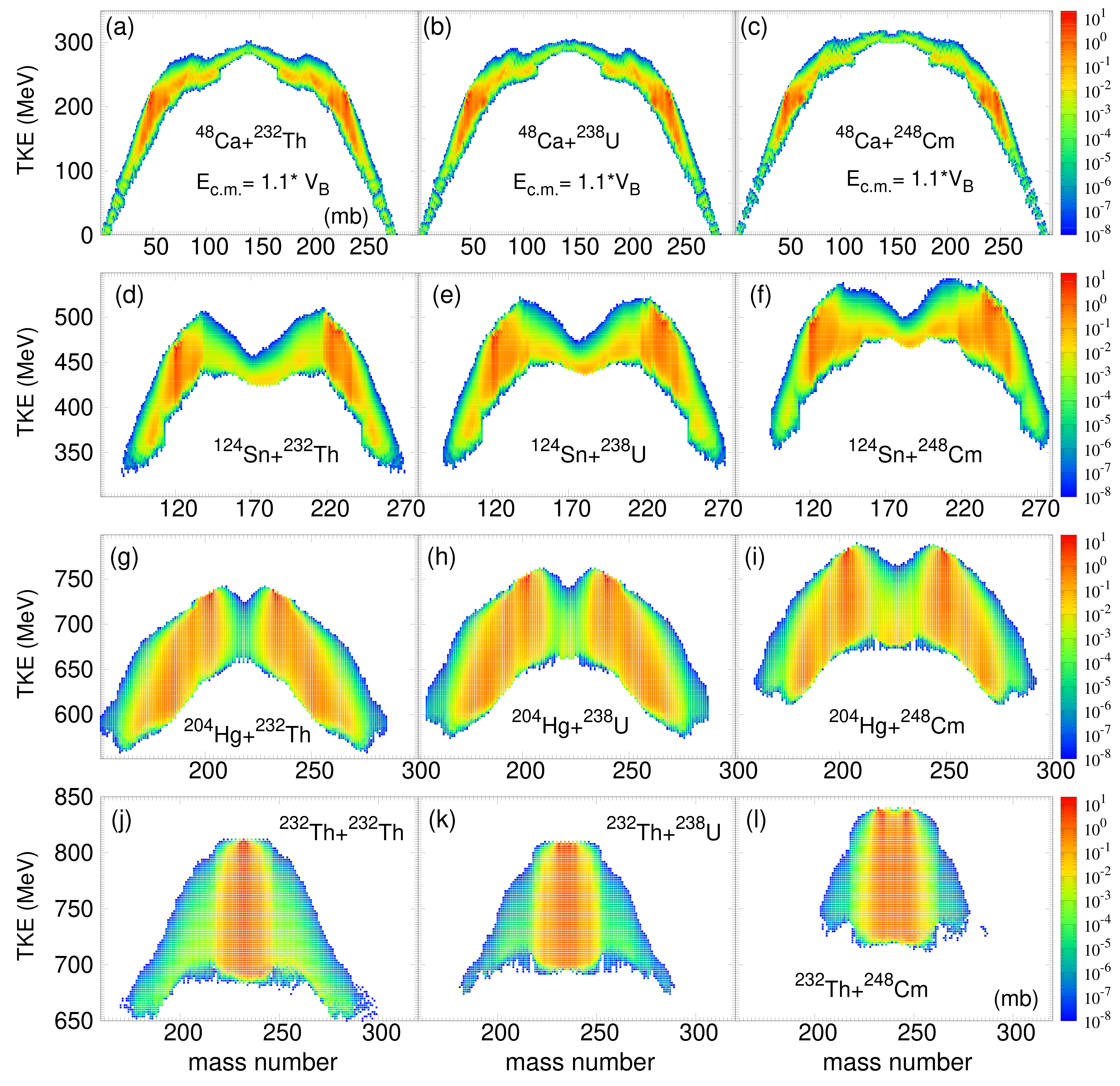}
\caption{\label{fig3}(Color online) Total kinetic energy (TKE) distribution of primary MNT products in the collisions of $^{48}$Ca, $^{124}$Sn, $^{204}$Hg and $^{232}$Th bombardment on $^{232}$Th, $^{238}$U and $^{248}$Cm at $E_{\rm c.m.}$ = 1.1*$V_{\rm B}$. }
\end{figure}

Using the numerical solution to deal with the master equation, we obtain the primary fragments with production probability and excitation energy. The total kinetic energy (TKE) of DNS is listed as a function of the mass number of all formed fragments.
Figture \ref{fig3} shows the calculation of TKE-mass distributions of the MNT primary products for $^{48}$Ca, $^{124}$Sn, $^{204}$Hg, $^{232}$Th induced MNT on targets $^{232}$Th, $^{238}$U and $^{248}$Cm with different mass asymmetry at incident energy $E_{\rm c.m.}$ = $1.1 \times V_{\rm B}$. This calculated TKE - mass of $^{48}$Ca + $^{248}$Cm agrees roughly well with available experimental data \cite{Zagrebaev_2005}.
The TKE-mass distribution is highly dependent on driving potential and mass asymmetry. Cross sections from TKE-mass distributions prefer to populate in pockets of driving potential. For the same target, the projectile is heavier and the TKE-mass distribution is wider. For the same projectile, targets with small mass variation have limited influence on the shape of TKE distribution. 
From the TKE-mass distributions, it was found that the tendency will be for mass to flow toward the trans-target fragments.
The shapes of TKE-mass show strong relevance with the impact parameter, which corresponds to sticking time of DNS.
For damped collisions, the sticking time is quite short (about units of 10$^{-21}$ s). These fast events correspond to grazing collisions with intermediate impact parameters, which are shown by the areas around injection points. A large amount of kinetic energy is dissipated very fast at relatively low mass transfer (more than 45 MeV during several units of 10$^{-21}$ s). The other events correspond to much slow collision with a large overlap of nuclear surface and significant nucleons rearrangement.

\begin{figure*}[htb]
\includegraphics[width=.9\linewidth]{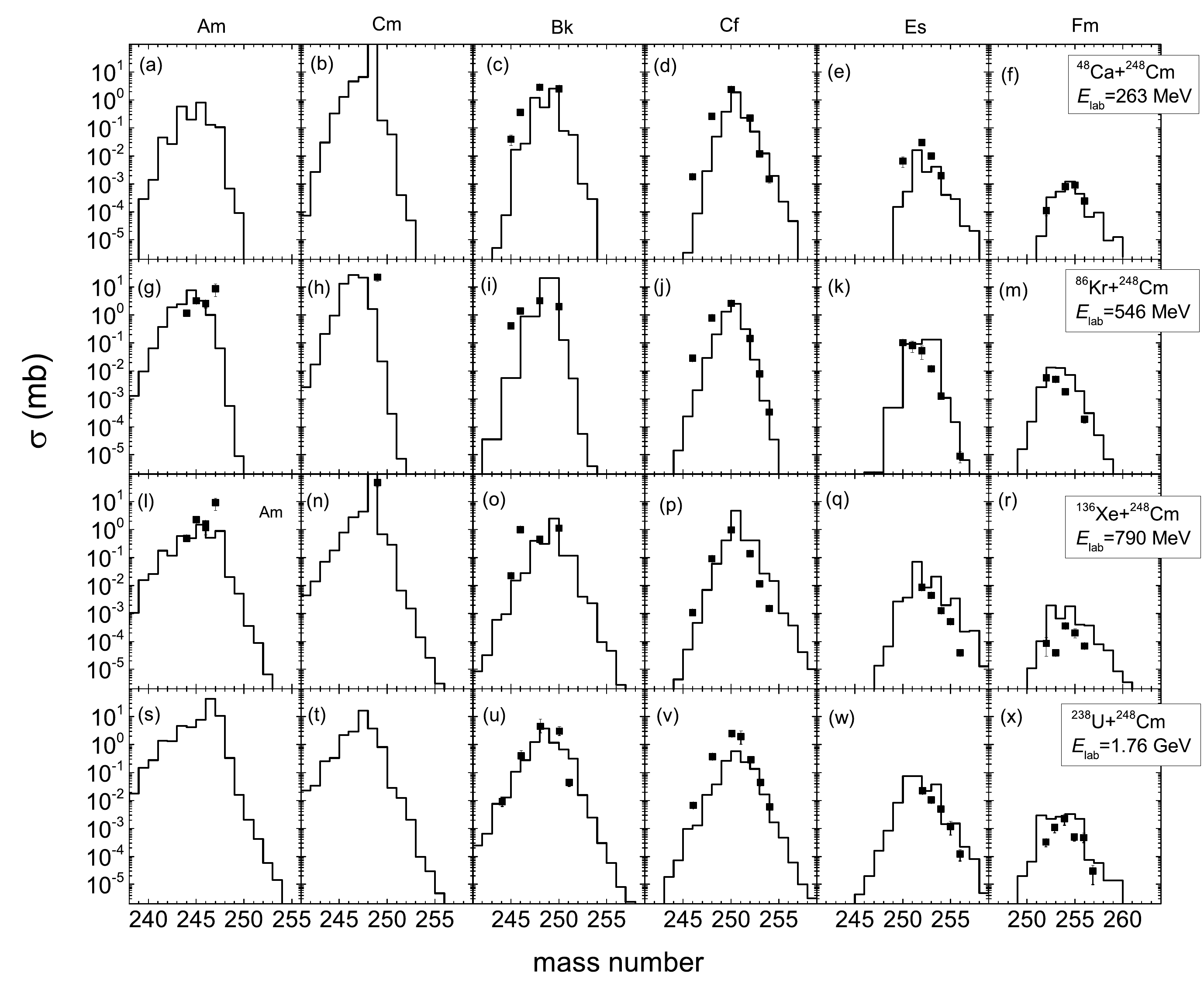}
\caption{\label{fig4}(Color online) 
Predicted secondary cross-sections of actinides in the reaction of $^{48}$Ca + $^{248}$Cm at $E_{\rm lab} = 263$ MeV, $^{86}$Kr + $^{248}$Cm at $E_{\rm lab} = 546$ MeV, $^{136}$Xe + $^{248}$Cm at $E_{\rm lab} = 790$ MeV and $^{238}$U + $^{248}$Cm at $E_{\rm lab} = 1760$ MeV (solid lines) compared to experimental data\cite{PhysRevC.31.1763,PhysRevC.33.1315,PhysRevC.88.054615} (solid square symbols wiht error bars)}
\end{figure*}

Predicted and experimental isotopic cross-sections of Am, Cm, Bk, Cf, Es and Fm  in $^{48}$Ca, $^{136}$Xe, $^{204}$Hg and $^{232}$Th induced reactions on $^{248}$Cm are shown in Fig. \ref{fig4}. 
The experimental data have been obtained at Lawrence Berkeley Laboratory (LBL) and Gesellschaft f$\ddot{\rm u}$r Schwerionenforschung (GSI) in the 1980s, which correspond to solid squares with error bars. the solid step lines are calculation results that were listed with experimental data used to make a comparison.
Our calculations are nicely consistent with the available experimental data of $^{48}$Ca + $^{248}$Cm at $E_{\rm lab} = 263$ MeV \cite{PhysRevC.33.1315}, of $^{86}$Kr + $^{248}$Cm at $E_{\rm lab} = 546$ MeV \cite{PhysRevC.33.1315}, of $^{136}$Xe + $^{248}$Cm at $E_{\rm lab} = 790$ MeV \cite{PhysRevC.31.1763}, of $^{238}$U + $^{248}$Cm at $E_{\rm lab} = 1760$ MeV \cite{PhysRevC.88.054615}. We found that collisions with lighter heavy-ions showed production cross-sections for the target-like actinides which were at least as large as those obtained using the very heavy projectiles $^{238}$U.
Strong enhancement in the production of Cf, Es, and Fm isotopes were produced when $^{238}$U were applied to bombard $^{248}$Cm as compared to $^{136}$Xe projectiles.
Reactions of $^{238}$U + $^{248}$Cm were used to explore the probability of synthesis superheavy nuclei. Due to the limitation of detection technology, only these events of Md, Fm, Es, Cf, and Bk isotopes have been obtained, the cross-sections have four orders of magnitude higher than in the $^{238}$U + $^{238}$U reactions.
Targets of $^{248}$Cm have been irradiated by beams of $^{136}$Xe and $^{86}$Kr at energies around the Coulomb barrier in the LBL's SuperHILAC. 
Production cross-sections and excitation functions of Md, Fm, Es, Cf, Bk, Cm, Am, Pu, Np, U, Pa isotopes
have been obtained, in which neutron-rich actinides production and neutron shell effect (N = 82 or Z = 50) have been discussed \cite{PhysRevC.33.1315}.

In reactions of $^{48}$Ca+$^{248}$Cm near Coulomb barrier energies, the TLFs and PLFs are dominant in all isotopic yields.
In the 1980s, to explore the effect of neutron-rich $^{48}$Ca on contributing to the yields of neutron-rich heavy actinides and to determine what effect the eight fewer neutrons in $^{40}$Ca have on the mass distribution, two series of experiments were performed at LBL and GSI\cite{PhysRevC.31.1763}.
Target-like fragments from Bk to Fm, and Rn, Ra, Ac, Th, U, Pu have been observed in the reactions of $^{48}$Ca + $^{248}$Cm at incident energies $E_{\rm lab}$ = 223-239, 248-263, 247-263, 272-288, 304-318 MeV.
In the year 2000, reactions of $^{48}$Ca+$^{248}$Cm at $E_{\rm lab}$ = 265.4, 270.2 MeV have been performed in GSI\cite{dev19}. In the experiment, MNT reaction products have been measured on SHIP. Due to the short detection time, a few transtarget isotopes have been obtained. They are $^{252,254}$ Cf, $^{254,256}$Es, $^{254,256}$Fm with available cross-sections. However, in our calculations, we could not reproduce the production cross-sections of very neutron-deficient isotopes. 

\begin{figure}[htb]
\includegraphics[width=1.\linewidth]{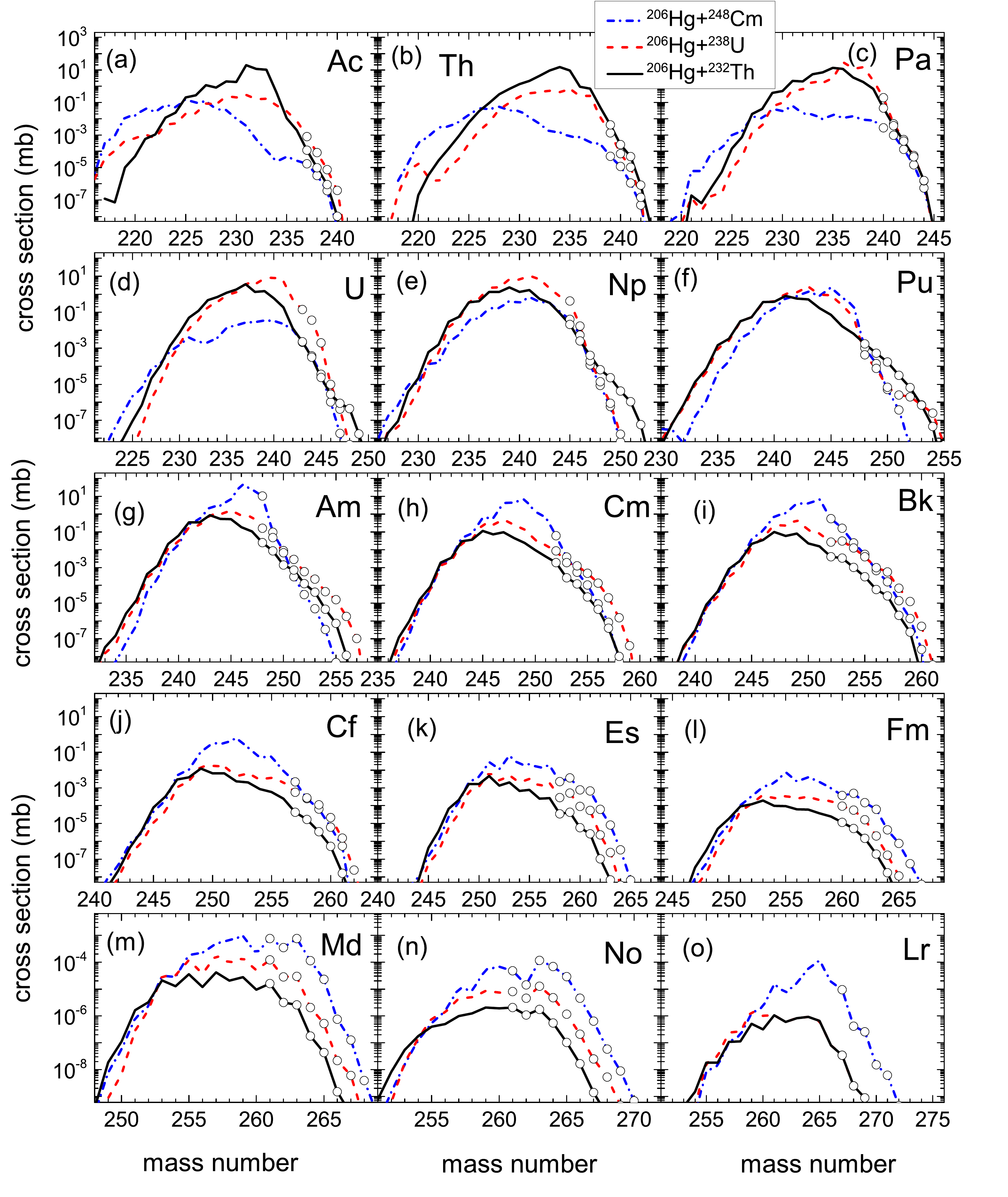}
\caption{\label{fig5}(Color online) Predicted secondary production cross-section of target-like fragments with Z = 89-103 in the collisions of $^{206}$Hg bombardment on $^{232}$Th, $^{238}$U and $^{248}$Cm, indicated by solid black, dash red and dash-dot lines, respectively. Unknown actinides are shown as open circles.}
\end{figure}

We calculate production cross-sections of actinides isotopes with Z = 89 - 103 systematically in targets $^{232}$Th-, $^{238}$U- and $^{248}$Cm-based reactions with projectiles from Ar to U. 
It was found that $^{204,206}$Hg as the projectile is favorable to produce unknown-yet neutron-rich actinide isotopes. To investigate the dependence of actinides production on targets, production of actinides isotopes with Z = 89 - 103 in reactions of $^{204,206}$Hg bombard on $^{232}$Th, $^{238}$U and $^{248}$Cm have been calculated systematically, correspond to solid black, red, blue lines respectively. Production of unknown neutron-rich actinide isotopes marked by open circles were added in Fig. \ref{fig5}. 

From Fig. \ref{fig5}, it was found that all three $^{232}$Th-, $^{238}$U- and $^{248}$Cm-based reactions could produce available cross-sections of new actinide isotopes. We mark $^{232}$Th-, $^{238}$U- and $^{248}$Cm-based reactions as R1, R2 and R3, respectively. 
For actinium, R2 prefer to produce unknown-yet $^{237,238,239,240}$Ac as 810 nb, 85 nb, 8 nb, 0.4 nb. 
For thorium, R1 prefer to produce unknown-yet $^{239,240,241,242}$Th as 4140 nb, 262 nb, 47 nb, 0.8 nb. 
For protactinium, R2 prefer to produce unknown-yet $^{240,241,242,243,244}$Pa as 0.2 mb, 4870 nb, 533 nb, 55 nb, 1.5 nb. 
For uranium, R1 and R2 prefer to produce unknown-yet $^{243,244,245,246}$U as 0.14 mb, 36770 nb, 2030 nb, 10 nb in R2, $^{247,248}$U as 0.8 nb, 0.4 nb for R1.
For neptunium, R1 and R2 prefer to produce unknown-yet $^{245,246,247}$Np as 0.4 mb, 17110 nb, 198 nb in R2, $^{249,250,251,252,253}$Np as 65 nb, 23 nb, 4 nb, 1 nb, 0.06 nb in R1.
For plutonium, R1 and R2 prefer to produce unknown-yet $^{248,249,250,251,252,253}$Pu as 1760 nb, 526 nb, 174 nb, 31 nb, 7 nb, 0.7 nb in R1, $^{254}$Pu as 0.3 nb in R2.
For americium, R3 and R2 prefer to produce unknown-yet $^{248,249,250}$Am as 10 mb, 0.1 mb, 9020 nb in R3, $^{251,252,253,254,255,256,257}$Am as 2850 nb, 585 nb, 225 nb, 43 nb, 16 nb, 2 nb, 0.1 nb in R2.
For curium, R2 and R3 prefer to produce unknown-yet $^{252,253}$Cm as 61450 nb, 3840 nb in R3, $^{254,255,256,257,258,259}$Cm as 1270 nb, 475 nb, 134 nb, 20 nb, 1.5 nb, 0.04 nb in R2.
For berkelium, R2 and R3 prefer to produce unknown-yet $^{252,253,254,255}$Bk as 0.55 mb, 0.16 mb, 24490 nb, 5520 nb in R3, $^{256,257,258,259,260}$Bk as 0.2 mb, 985 nb, 572 nb, 70 nb, 12 nb, 0.5 nb in R2.
For californium, R3 and R2 prefer to produce unknown-yet $^{257,258}$Cf as 2180 nb, 245 nb in R3, $^{259,260,261,262}$Cf as 114 nb, 21 nb, 1.5 nb, 0.02 nb in R2.
For einsteinium, R3 prefer to produce unknown-yet $^{258,259,260,261,262,263,264}$Es as 2220 nb, 3640 nb, 759 nb, 675 nb, 63 nb, 8 nb, 0.3 nb.
For fermium, R3 prefer to produce unknown-yet $^{260,261,262,263,264,265}$Fm as 372 nb, 493 nb, 147 nb, 63 nb, 4.4 nb, 0.7 nb.
For mendelevium, R3 prefer to produce unknown-yet $^{261,262,263,264,265,266}$Md as 756 nb, 348 nb, 763 nb, 113 nb, 23 nb, 0.7 nb.
For nobelium, R3 prefer to produce unknown-yet $^{261,262,263,264,265,266}$No as 47 nb, 14 nb, 11 nb, 7 nb, 3 nb, 4.5 nb, 0.6 nb.
For lawrencium, R3 prefer to produce unknown-yet $^{267,268,269,270}$Lr as 9 nb, 410 pb, 254 pb, 15 pb.
It should be noticed that unknown actinide products are highly dependent on target mass. 
They could not reach the unknown proton-rich actinides region.

\begin{figure}[htb]
\includegraphics[width=1.\linewidth]{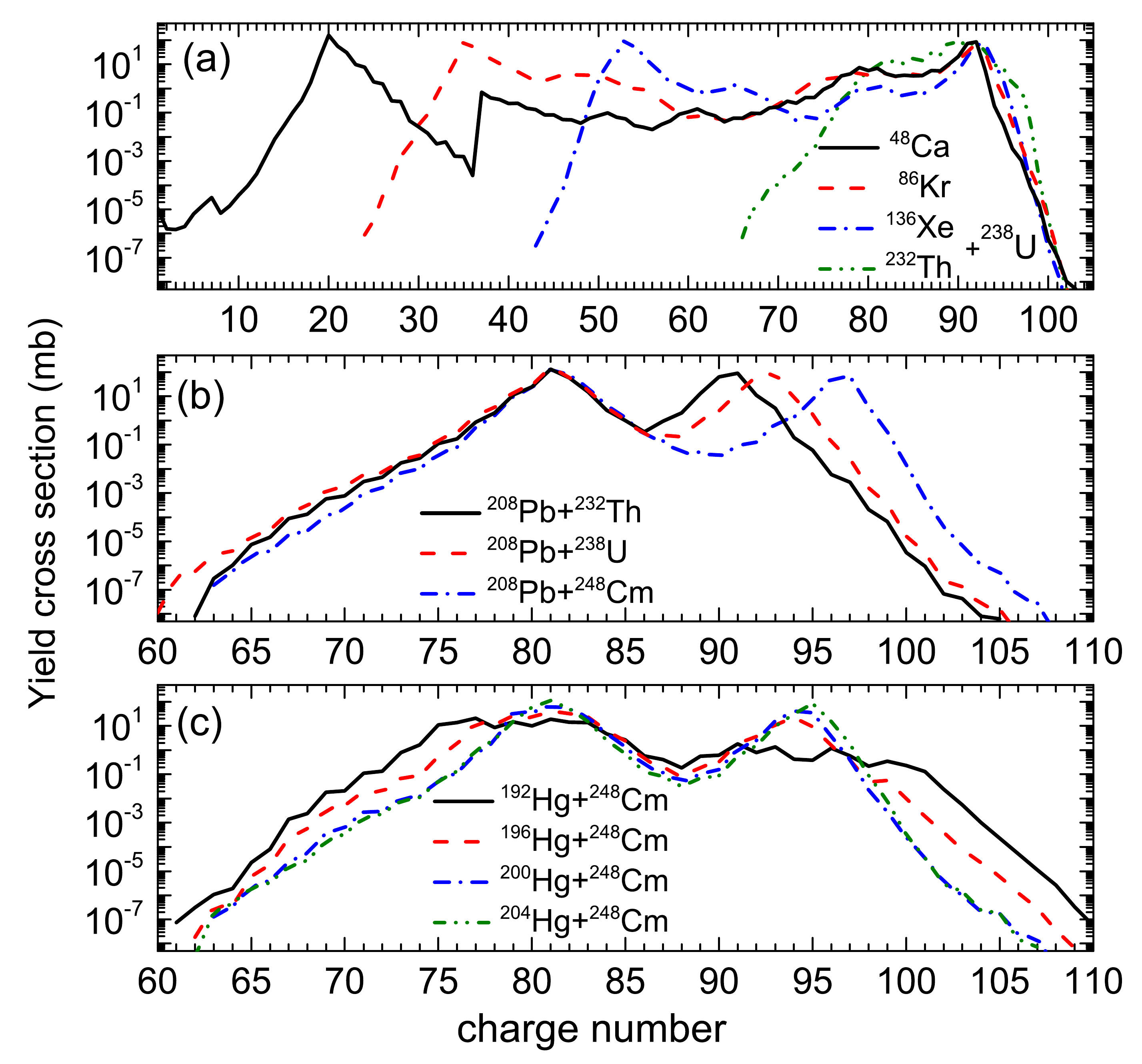}
\caption{\label{fig6}(Color online) 
Yields-charge distributions of survived fragments formation in the $^{48}$Ca, $^{86}$Kr, $^{136}$Xe, $^{192,196,200,204}$Hg, $^{208}$Pb and $^{232}$Th induced reactions with $^{232}$Th, $^{238}$U and $^{248}$Cm at incident energies $E_{\rm c.m.}$ = 1.1*$V_{\rm B}$.
In panel (a), solid black, dash red, dash-dot blue and dash-dot-dot olive lines are for $^{48}$Ca, $^{86}$Kr, $^{136}$Xe and $^{232}$Th induced reactions with $^{238}$U, respectively. In panel (b), solid black, dash red, dash-dot blue lines stand for $^{208}$Pb induced reactions on different targets $^{232}$Th, $^{238}$U and $^{248}$Cm, respectively. In panel (c), solid black, dash red, dash-dot blue and dash-dot-dot olive lines represent isotopes $^{192,196,200,204}$Hg induced collisions with $^{248}$Cm, respectively.}
\end{figure}

Dependence of charge distribution of survived actinide production yields on the projectile mass (isospin) with a given target and target mass with given projectiles have been shown in Fig. \ref{fig6}.
In panel (a), $^{48}$Ca, $^{86}$Kr, $^{136}$Xe, $^{232}$Th bombard on $^{238}$U are represent by solid black, red, blue and olive lines, respectively.
It was found that two peaks of charge distribution of MNT products are around the projectile-target positions. MNT products in lighter projectiles-induced reactions could cover a wider charge number region. The heaviest projectile-induced MNT seems to produce the largest charge distribution yields for target-like elements.
In panel (b), $^{208}$Pb induced MNT reactions with $^{232}$Th, $^{238}$U and $^{248}$Cm are shown by solid black, red and blue lines.
It was found that $^{248}$Cm-based reactions prefer to produce actinide isotopes with $Z \ge $95 at larger cross-sections.
$^{238}$U-based reactions prefer to produce actinide isotopes with 92 $\le Z < $95 at larger cross-sections.
$^{232}$Th-based reactions prefer to produce actinide isotopes with  89 $\le Z < $92 at larger cross-sections.
In panel (c), isotopes $^{192,196,200,204}$Hg induced MNT reactions with $^{248}$Cm are shown by solid black, red and blue lines.
It should be noticed that neutron-deficient projectile ${192}$Hg-induced reactions do not have an obvious peak around injection points, because the driving potential at its injection point is off the minimum valley trajectory, however, which is contribute to diffuse far beyond the target, even the superheavy nuclei region. 
\begin{figure}[htb]
\includegraphics[width=1.\linewidth]{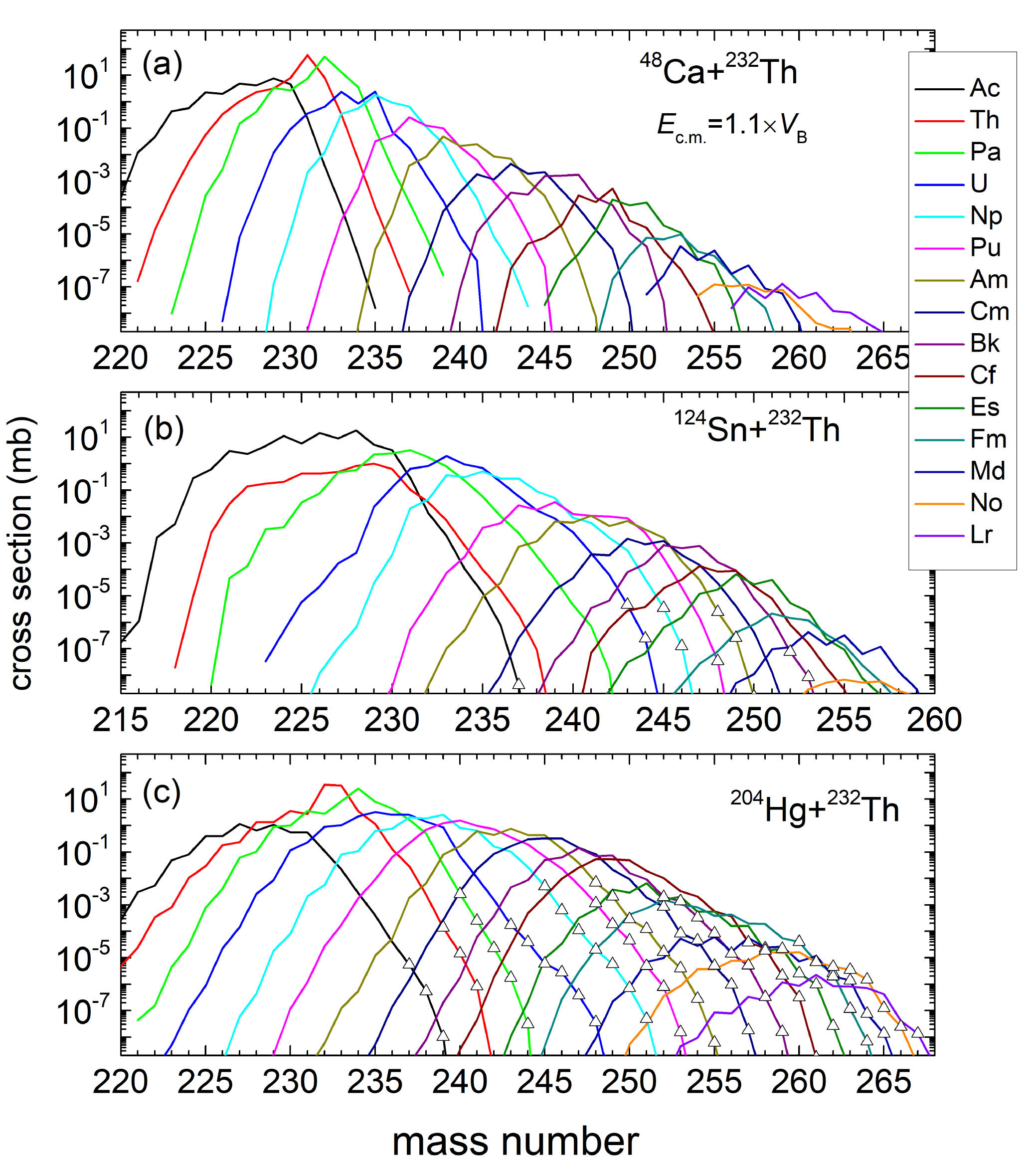}
\caption{\label{fig7}(Color online) In the above panels, solid black, red, green, blue, cyan, magenta, dark yellow, navy, purple, wine, olive, dark cyan, royal, orange, violet lines indicate survived Actinium, Thorium, Protactinium, Uranium, Neptunium, Plutonium, Americium, Curium, Berkelium, Californium, Einsteinium, Fermium, Mendelevium, Nobelium, Lawrencium isotopic cross-section distribution, respectively. Isotopes yields in panel (a), (b), (c) correspond to $^{48}$Ca, $^{124}$Sn and $^{204}$Hg induced reaction with $^{232}$Th at incident energy $E_{\rm c.m.}$ = 1.1*$V_{\rm B}$, respectively. Open-up triangles stand for unknown actinide isotopes.}
\end{figure}

Dependence of survived actinide production cross-section from Z = 89 - 103 in $^{48}$Ca, $^{124}$Sn and $^{204}$Hg induced reaction with $^{232}$Th at $E_{\rm c.m.}$ = 1.1*$V_{\rm B}$ were shown in Fig. \ref{fig7}. 
All actinide isotopic distributions from Ac to Lr are marked by solid black, red, green, blue, cyan, magenta, dark yellow, navy, purple, wine, olive, dark cyan, royal, orange, and violet lines in order.
It was found that secondary actinide production is highly dependent on projectile mass, especially for projectile mass in a great shift. 
The more heavy projectile-induced MNT reaction seems to produce more new neutron-rich actinide isotopes, which consist of products in $^{206}$Hg induced reactions with $^{248}$Cm.

\begin{figure*}[htb]
\includegraphics[width=.9\linewidth]{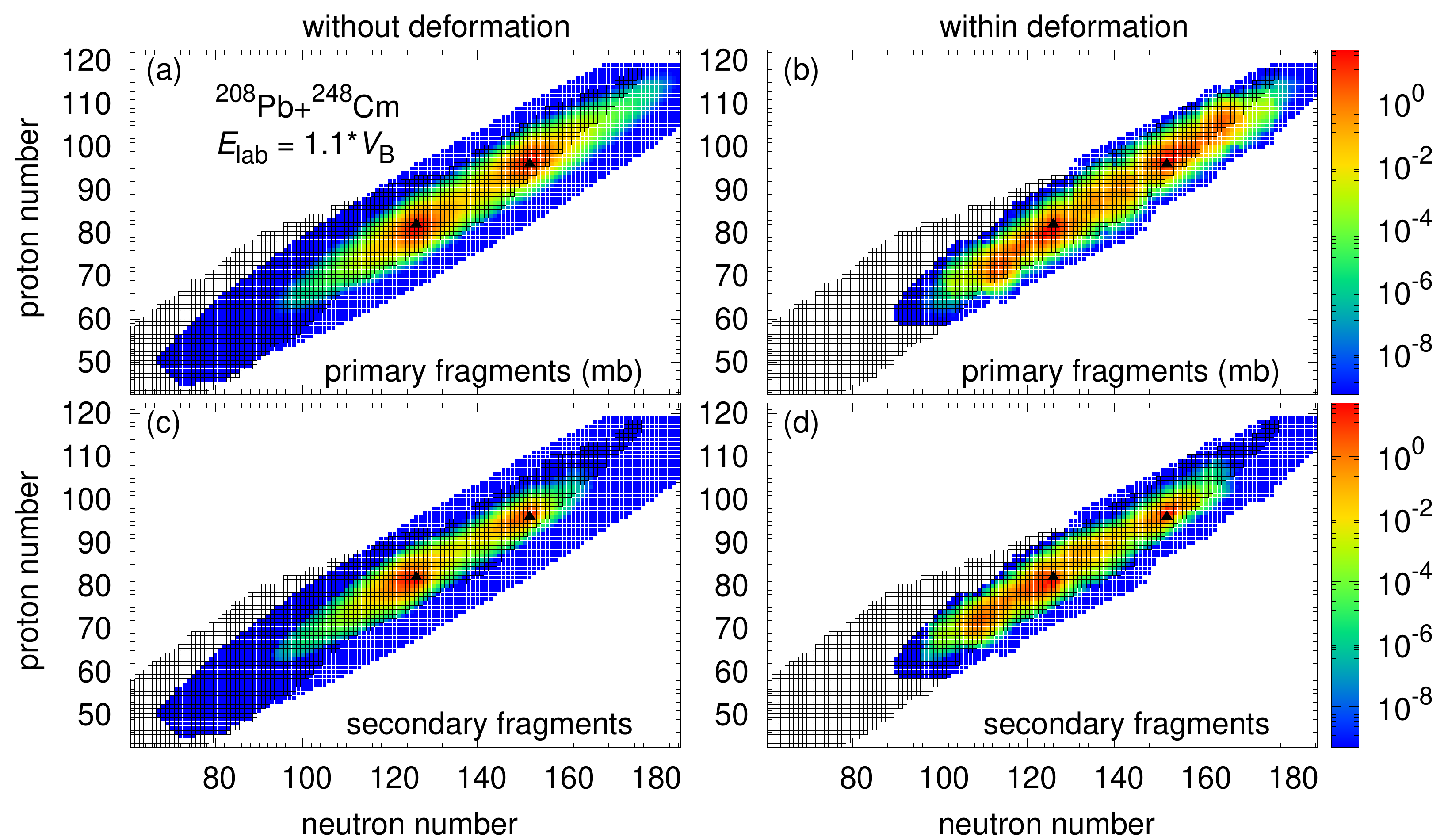}
\caption{\label{fig8}(Color online) 
For collisions of $^{208}$Pb + $^{248}$Cm at $E_{\rm c.m.}$ = 1.1*$V_{\rm B}$, the calculation cross-sections of primary and secondary fragments with and without dynamical deformation were shown in panel (a), (b), (c), (d), respectively. Solid triangles stand for projectile-target injection points.}
\end{figure*}

Figure \ref{fig8} shows comparison of calculations for primary and secondary yields of $^{208}$Pb + $^{248}$Cm at $E_{\rm c.m.}$ = 1.1*$V_{\rm B}$ within and without quadrupole deformation. Open black squares stand for existing isotopes in the nuclide chart, which are added with calculation to highlight the unknown isotopes. Solid black triangles represent the injection points. Compare to the calculation involving deformation, calculation without deformation tends to the unknown neutron-rich region easily, which predicts more new isotopes. Yield distribution without deformation is more smooth than that with dynamical deformation. 
From Fig. \ref{fig8}, it was found that primary yields cover the whole actinide region as very large formation cross-sections. The de-excitation process depressed the formation of secondary cross-section at four orders of magnitude level through fission, evaporating light particles, and gamma-ray.

\begin{figure*}[htb]
\includegraphics[width=1.\linewidth]{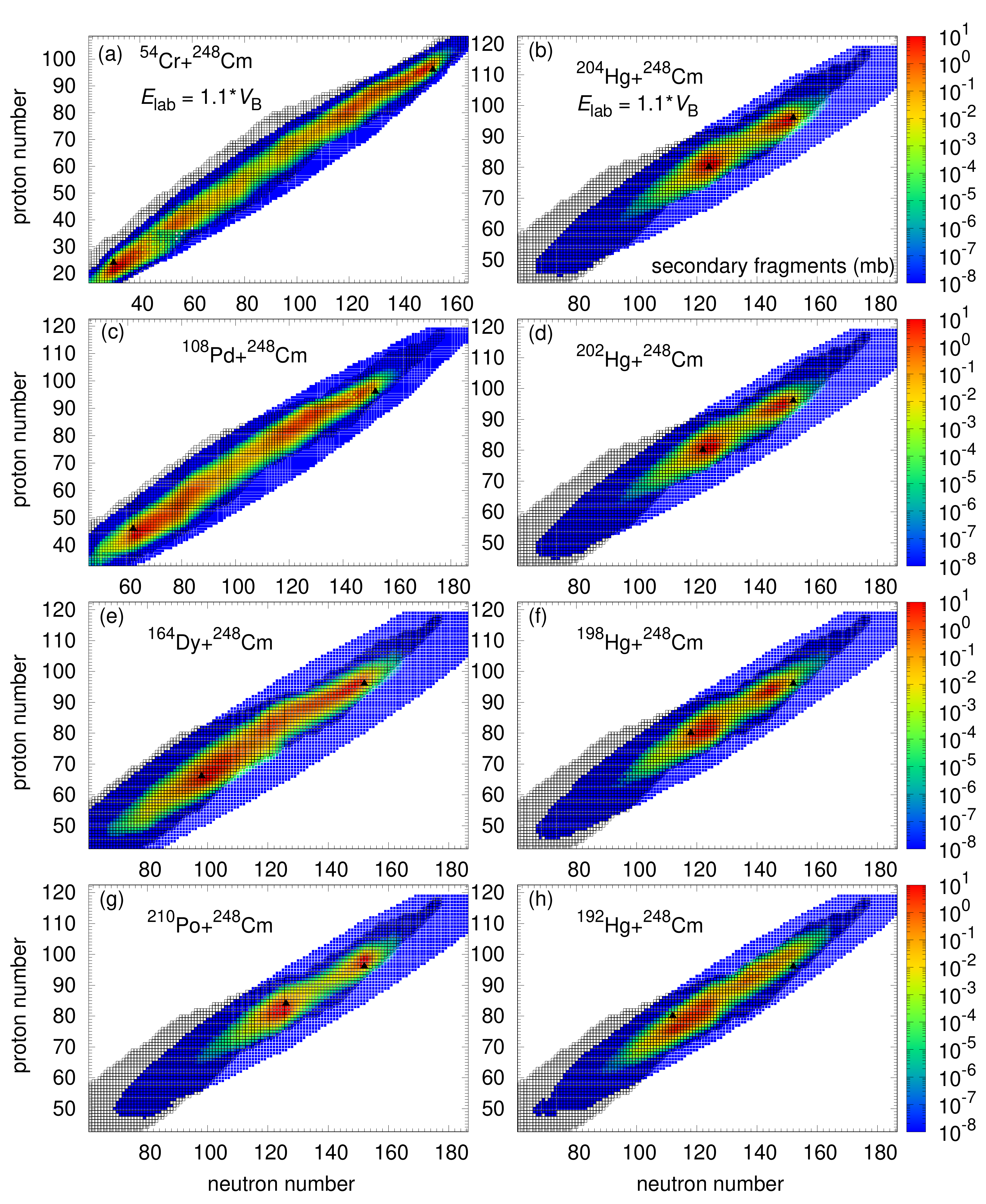}
\caption{\label{fig9}(Color online) The calculated yields of secondary fragments in collisions of $^{54}$Cr, $^{108}$Pd, $^{164}$Dy, $^{210}$Po, $^{192,198,202,204}$Hg on $^{248}$Cm at $E_{\rm c.m.}$ = 1.1*$V_{\rm B}$ were shown in (a), (b), (c), (d), (e), (f), (g) and (h), respectively. Solid black triangles stand for projectile-target injection points.}
\end{figure*}

Yields distributions in target $^{248}$Cm-based reactions with mass asymmetries and different isospin projectiles were shown in Fig. \ref{fig9}.
From panels (a) and (c), it was found that projectiles with lighter mass-induced reactions diffuse more broadly, compared to heavy projectiles induced reactions.
In panels (b), (d), (f), and (h), yields of more neutron-rich projectiles induced reactions tend to shift to the unknown neutron-rich side, especially in the actinide region. 
In panels (a), (c), (e), and (g), yields of heavier mass projectiles-induced reactions tend to shift to the unknown neutron-rich side, especially in the actinide region easily. 
For $^{164}$Dy + $^{248}$Cm at $E_{\rm c.m.} = 1.1\times V_{\rm B}$, production cross-sections of neutron-rich fragments with $Z=60-80$ have been predicted magnitude level around hundreds nanobarns.

\section{Conclusions}\label{sec4}

In summary, based on DNS model, we perform calculations for targets $^{232}$Th-, $^{238}$U- and $^{248}$Cm-based reactions with projectiles from over the nuclide chart at incident energies $E_{\rm c.m.}$ = 1.1*$V_{\rm B}$, systematically. 
The formation probabilities of MNT fragments diffuse along PES, resulting in the occupation of the whole nuclide chart, especially covering the nuclei drip lines and unknown superheavy region. However, highly exciting fragments are rapidly fissile.  Survival probabilities for these fragments in the unknown nuclide chart are quite low. 
Our calculations of $^{48}$Ca, $^{86}$Kr, $^{136}$Xe and $^{238}$U induced reactions with $^{248}$Cm at incident energies around Coulomb barriers have a good agreement with available experimental data reasonably.
In this work, PES and TKE-mass distributions for selected reactions were exported from our calculations, which help us to predict the production cross-section trends.
The influence of mass asymmetry and isospin effect on actinide products in these MNT reactions have been investigated thoroughly. 
In our calculations, MNT products are highly dependent on mass asymmetry. Heavier projectiles induced MNT products gave wider isotopic chains and larger cross-sections, showing a greater probability of synthesis of unknown actinide around the drip line. 
Dependence of actinide products on projectiles isospin effect has been studied by Ca, Xe, Sn, Xe, and Hg isotopes induced MNT reactions with targets $^{232}$Th, $^{238}$U and $^{248}$Cm.
It was found that the neutron-rich projectiles such as $^{204,206}$Hg enhanced the production of the neutron-rich actinides. 
A projectile-target combination of an actinide-based target and a heavy neutron-rich projectile such as $^{204,206}$Hg + $^{248}$Cm were suggested to produce new neutron-rich actinide isotopes, and even could be used to produce new superheavy nuclei. In this work, formation cross-sections of massive unknown actinides with Z = 89-103 nearby drip lines were predicted as the level magnitude from nanobarn to millibarn.

\section{Acknowledgements}
This work was supported by the National Science Foundation of China (NSFC) (Grants No. 12105241,12175072), NSF of Jiangsu Province (Grants No. BK20210788), Jiangsu Provincial Double-Innovation Doctor Program(Grants No. JSSCBS20211013) and University Science Research Project of Jiangsu Province (Grants No. 21KJB140026). This project was funded by the Key Laboratory of High Precision Nuclear Spectroscopy, Institute of Modern Physics, Chinese Academy of Sciences (IMPKFKT2021001).

%

\end{document}